\documentclass[10.75pt,preprint]{article}
\usepackage[english]{babel}
\usepackage{amsmath,amssymb,graphicx,calc,capt-of,ifthen,theorem,pifont,amsfonts}

\begin{document}

\title{Non-Relativistic $\tilde{\delta}$ Gravity: A Description of Dark Matter.}
\author{Jorge Alfaro\footnote{Facultad de F\'isica, Pontificia Universidad Cat\'olica de Chile. Casilla 306, Santiago, Chile. jalfaro@uc.cl.} and Pablo Gonz\'alez\footnote{Departamento de F\'isica, FCFM, Universidad de Chile. Blanco Encalada 2008, Santiago, Chile. pgonzalez@ing.uchile.cl (Current Affiliation).} \footnote{Facultad de F\'isica, Pontificia Universidad Cat\'olica de Chile. Casilla 306, Santiago, Chile. pegonza2@uc.cl.}}
\maketitle

\begin{abstract}
$\tilde{\delta}$ Gravity is a gravitational field model, where the geometry is governed by two symmetric tensors, $g_{\mu \nu}$ and $\tilde{g}_{\mu \nu}$, and new matter fields ($\tilde{\delta}$ Matter fields) are added to the original matter fields. These new components appear motivated by a new symmetry, called $\tilde{\delta}$ symmetry. In previous works, the model is used to explain the expansion of the Universe without Dark Energy. This result and the additional contribution to the mass by $\tilde{\delta}$ Matter are motivations to study the Dark Matter effect with $\tilde{\delta}$ Gravity. In this work, we will derive the Non-Relativistic limit to obtain a correction to the rotation velocity in a galaxy, then we will analyze the most common galaxy density profiles to describe Dark Matter.
\end{abstract}

\section*{Introduction.}

Recent discoveries in cosmology have revealed that most part of matter is in the form of unknown matter, Dark Matter \cite{DM DE 1}-\cite{DM New 8}, and that the dynamics of the expansion of the Universe is governed by a mysterious component that accelerates its expansion, the so called Dark Energy \cite{DM DE 2}-\cite{DM DE 4}. That is the Dark Sector. Although General Relativity (GR) is able to accommodate the Dark Sector, its interpretation in terms of fundamental theories of elementary particles is problematic \cite{DM DE 5}. In relation to Dark Energy, the accelerated expansion of the universe can be explained if a small cosmological constant ($\Lambda$) is present. At early times, this constant is irrelevant, but at the later stages of the evolution of the Universe $\Lambda$ will dominate the expansion, explaining the observed acceleration. However $\Lambda$ is too small to be generated in quantum field theory (QFT) models, because $\Lambda$ is the vacuum energy, which is usually predicted to be very large \cite{lambda problem}.\\

On the other side, some candidates exist that could play the role of Dark Matter. Many experiments are been carried out to determine its nature \cite{DM DE 1}, however its detection has been problematic. In a galactic scale, Dark Matter produces an anomalous rotation velocity where it is relatively constant far from the center of the galaxy \cite{DM New 1}-\cite{DM New 8}. Usually, some density profiles are used to represent the matter distribution in a galaxy, including the Dark matter effect \cite{Einasto 1}-\cite{NFW 3}. On the other side, a lot of alternative models, where a modification to gravity is introduced, have been developed to explain the Dark Matter effect. For instance, a explanation based on the modification of the dynamics for small accelerations cannot be ruled out \cite{DM DE 6,DM DE 7}.\\

Recently, in \cite{delta gravity}, a model of gravitation, very similar to GR, is presented. In that paper, we considered two different points. The first is that GR is finite on shell at one loop in vacuum {\cite{tHooft}}, so renormalization is not necessary at this level. The second is the $\tilde{\delta}$ gauge theories (DGT) originally presented in {\cite{Alfaro 2,Alfaro 3}}, where the main properties are: (a) A new kind of field $\tilde{\phi}_{I}$ is introduced, different from the original set $\phi_{I}$. (b) The classical equations of motion of $\phi_{I}$ are satisfied even in the full quantum theory. (c) The model lives at one loop. (d) The action is obtained through the extension of the original gauge symmetry of the model, introducing an extra symmetry that we call $\tilde{\delta}$ symmetry, since it is formally obtained as the variation of the original symmetry. When we apply this prescription to GR, we obtain $\tilde{\delta}$ Gravity. For these reasons, the original motivation was to develop the quantum properties of this model (See \cite{delta gravity}), but now we prefer to emphasize the use of $\tilde{\delta}$ Gravity as an effective model of gravitation and explore its phenomenological predictions.\\

In \cite{DG DE,Paper DE}, we presented a truncated version of $\tilde{\delta}$ Gravity applied to Cosmology. The $\tilde{\delta}$ symmetry was fixed in different ways in order to simplify the analysis of the model to explain the accelerated expansion of the universe without Dark Energy. The results were quite reasonable taking into account the simplifications involved, but $\tilde{\delta}$ Matter was ignored in the process. After in \cite{Paper 1}, we developed the Cosmological solution in a $\tilde{\delta}$ Gravity version where $\tilde{\delta}$ symmetry is preserved, which means that we are forced to include $\tilde{\delta}$ Matter. In that case, the accelerated expansion can be explained in the same way and additionally we have a new component of matter as Dark Matter candidate. Besides, we guaranteed that the special properties of $\tilde{\delta}$ Theories previously mentioned are preserved.\\

In this work, we will analyze the Dark Matter phenomenon to complete the Dark Sector in $\tilde{\delta}$ Gravity with $\tilde{\delta}$ Matter. Specifically, we will study the Non-Relativistic case to understand the behavior of the model in the (Post)-Newtonian limit. We have that $\tilde{\delta}$ Gravity agrees with GR at the classical level far from the sources. In particular, the causal structure of $\tilde{\delta}$ Gravity in vacuum is the same as in GR. However, inside a matter fluid like a galaxy, new effects appear because of $\tilde{\delta}$ Matter. In \textbf{Section \ref{Sec: delta Gravity and Non-Relativistic Limit}}, we will present the $\tilde{\delta}$ Gravity action that is invariant under extended general coordinate transformation. We will find the equations of motion of this action. We will see that the Einstein's equations continue to be valid and we will obtain a new equation for $\tilde{g}_{\mu \nu}$. In these equations, two energy momentum tensors, $T_{\mu \nu}$ and $\tilde{T}_{\mu \nu}$, are defined. Then, we will introduce the non-relativistic case, where we compute the Newtonian and Post-Newtonian limit and a relation between the ordinary density and $\tilde{\delta}$ Matter density is found. From this, we will obtain an effective density, giving us an additional mass due to $\tilde{\delta}$ Matter. This effective density produces a modification in the rotation velocity. In \textbf{Section \ref{Sec: Density Profiles}}, we will do an analysis to explain Dark Matter with $\tilde{\delta}$ Matter in a galactic scale, using the effective density found in the last section and some realistic density profiles as Einasto and Navarro-Frenk-White (NFW) profiles. At Newtonian level, $\tilde{\delta}$ Gravity is very similar to GR when $\tilde{\delta}$ Matter is negligible. We will see that $\tilde{\delta}$ Matter effect is not related to the scale, but rather to the behavior of the distribution of ordinary matter. In the solar system scale, where the large structures as planets and stars have a concentrated and almost constant distributions, $\tilde{\delta}$ Matter is practically negligible. However, in a galactic scale, where the distribution is strongly dynamic, $\tilde{\delta}$ Matter is important to explain Dark Matter. So, the Dark Matter effect could be explained with a considerably less quantity of ordinary Dark Matter, explaining its extremely problematic detection.\\

Before continuing, we want to introduce a word of caution. In what follows, we want to study $\tilde{\delta}$ Gravity as a \textit{classical effective model}. This means to approach the problem from the phenomenological side instead of neglecting it \textit{a priori} because it does not satisfy yet all the properties of a fundamental quantum theory. In a cosmological level, the observations indicate that a phantom component is compatible with most classical tests of cosmology based on current data \cite{phantom 1}-\cite{phantom 6}. The nature of the Dark Sector is such an important and difficult cosmological problem that cosmologists do not expect to find a fundamental solution of it in one stroke and are open to explore new possibilities. Now, the \textit{phantom problem} is being studied in this moment and the results will be presented in a future work.\\

Additionally, it should be remarked that $\tilde{\delta}$ Gravity is not a metric model of gravity because massive particles do not move on geodesics. Only massless particles move on null geodesics of a linear combination of both tensor fields. Additionally, it is important to notice that we will work with the $\tilde{\delta}$ modification for GR, based on the Einstein-Hilbert theory. From now on, we will refer to this model as $\tilde{\delta}$ Gravity.\\

\newpage

\section{\label{Sec: delta Gravity and Non-Relativistic Limit}$\tilde{\delta}$ Gravity and Non-Relativistic Limit.}

Using the prescription given in \textbf{Appendix A}, we will present the action of $\tilde{\delta}$ Gravity and then we will derive the equations of motion. Then, we will consider the Newtonian and Post-Newtonian limit to study new effects on $\tilde{\delta}$ Gravity. We expect a weak deviation of GR at solar system scale, but stronger in a galactic scale to understand the Dark Matter phenomenon like a $\tilde{\delta}$ Gravity effect. We will principally pay attention to $\tilde{\delta}$ Matter contribution.\\

\subsection{\label{SubSec: Equations of Motion}Equations of Motion:}

To obtain the action of $\tilde{\delta}$ Gravity, we will consider the Einstein-Hilbert Action:

\begin{eqnarray}
\label{EH action}
S_0 = \int d^4x \sqrt{-g} \left(\frac{R}{2\kappa} + L_M\right),
\end{eqnarray}

where $L_M = L_M(\phi_I,\partial_{\mu}\phi_I)$ is the lagrangian of the matter fields $\phi_I$. Using (\ref{Action}) from \textbf{Appendix A}, this action becomes:

\begin{eqnarray}
\label{grav action}
S = \int d^4x \sqrt{-g} \left(\frac{R}{2\kappa} + L_M - \frac{1}{2\kappa}\left(G^{\alpha \beta} - \kappa T^{\alpha \beta}\right)\tilde{g}_{\alpha \beta} + \tilde{L}_M\right),
\end{eqnarray}

where $\kappa = \frac{8 \pi G}{c^2}$, $\tilde{g}_{\mu \nu} = \tilde{\delta}g_{\mu \nu}$ and:

\begin{eqnarray}
\label{EM Tensor}
T^{\mu \nu} = \frac{2}{\sqrt{-g}} \frac{\delta}{\delta g_{\mu \nu}}\left[\sqrt{-g} L_M\right] \\
\label{tilde L matter}
\tilde{L}_M = \tilde{\phi}_I\frac{\delta L_M}{\delta \phi_I} + (\partial_{\mu}\tilde{\phi}_I)\frac{\delta L_M}{\delta (\partial_{\mu}\phi_I)},
\end{eqnarray}

where $\tilde{\phi}_I = \tilde{\delta}\phi_I$ are the $\tilde{\delta}$ Matter fields. These matter components are the fundamental difference with \cite{Paper DE}. From this action, we can see that the Einstein's equations are preserved and an additional equation for $\tilde{g}_{\mu \nu}$ is obtained. So, the equations of motion are:

\begin{eqnarray}
\label{Einst Eq} G^{\mu \nu} &=& \kappa T^{\mu \nu} \\
\label{tilde Eq} F^{(\mu \nu) (\alpha \beta) \rho
\lambda} D_{\rho} D_{\lambda} \tilde{g}_{\alpha \beta} + \frac{1}{2}g^{\mu \nu}R^{\alpha \beta}\tilde{g}_{\alpha \beta} - \frac{1}{2}\tilde{g}^{\mu \nu}R &=& \kappa\tilde{T}^{\mu \nu}.
\end{eqnarray}

with:

\begin{eqnarray}
\label{F}
F^{(\mu \nu) (\alpha \beta) \rho \lambda} &=& P^{((\rho
\mu) (\alpha \beta))}g^{\nu \lambda} + P^{((\rho \nu) (\alpha
\beta))}g^{\mu \lambda} - P^{((\mu \nu) (\alpha \beta))}g^{\rho
\lambda} - P^{((\rho \lambda) (\alpha \beta))}g^{\mu \nu} \nonumber \\
P^{((\alpha \beta)(\mu \nu))} &=& \frac{1}{4}\left(g^{\alpha
\mu}g^{\beta \nu} + g^{\alpha \nu}g^{\beta \mu} - g^{\alpha
\beta}g^{\mu \nu}\right),
\end{eqnarray}

where $(\mu \nu)$ denotes that $\mu$ and $\nu$ are in a totally symmetric combination. An important fact to notice is that our equations are of second order in derivatives which is needed to preserve causality. Finally, from \textbf{Appendix A}, we have that the action (\ref{grav action}) is invariant under (\ref{trans g}) and (\ref{trans gt}). This means that two conservation rules are satisfied. They are:

\begin{eqnarray}
\label{Conserv T}
D_{\nu}T^{\mu \nu} &=& 0 \\
\label{Conserv tilde T}
D_{\nu}\tilde{T}^{\mu \nu} &=& \frac{1}{2}T^{\alpha \beta}D^{\mu}\tilde{g}_{\alpha \beta} - \frac{1}{2}T^{\mu \beta} D_{\beta}\tilde{g}^{\alpha}_{\alpha} + D_{\beta}(\tilde{g}^{\beta}_{\alpha}T^{\alpha \mu}).
\end{eqnarray}

In conclusion, we have that the original equation in the Einstein-Hilbert Action, given by (\ref{Einst Eq}) and (\ref{Conserv T}), are preserved. This is one of the principal properties of $\tilde{\delta}$ Theories. Additionally, we have two new equations, (\ref{tilde Eq}) and (\ref{Conserv tilde T}), to obtain the solution of $\tilde{g}_{\mu \nu}$ and $\tilde{\delta}$ Matter respectively. These components will produce new contributions to some phenomena, usually taken into account in (\ref{Einst Eq}) and (\ref{Conserv T}) in GR. For example, in \cite{Paper 1}, a universe without a cosmological constant is considered, so the accelerated expansion of the universe is produced by the additional effect of $\tilde{\delta}$ components. In the same way, in this paper, we will see if $\tilde{\delta}$ Gravity give us a new contribution to Dark Matter. To do this, we will study the Post-Newtonian limit.\\

\subsection{\label{SubSec: Newtonian and Post-Newtonian limit}Newtonian and Post-Newtonian limit:}

If we introduce one order more for the Newtonian limit, the metric will be given by \cite{Weinberg grav}:

\begin{eqnarray}
\label{g PostNewton} g_{\mu \nu}dx^{\mu}dx^{\nu} &=&  - \left(1+2\phi\epsilon^2+2\left(\phi^2 + \psi\right)\epsilon^4\right) \left(\frac{c dt}{\epsilon}\right)^2 + \left(1 - 2\phi\epsilon^2 - 2\psi\epsilon^4\right)\left(dx^2+dy^2+dz^2\right) \nonumber \\
&& + 2\epsilon^3\left(\chi_1dx + \chi_2dy + \chi_3dz\right) \left(\frac{c dt}{\epsilon}\right) + \epsilon^4\left(\xi_{11}dx^2 + \xi_{22}dy^2 + \xi_{33}dz^2 \right. \nonumber \\
&& \left.+ 2\xi_{12} dx dy + 2\xi_{13} dx dz + 2\xi_{23} dy dz\right),
\end{eqnarray}

where $\epsilon \sim \frac{v}{c}$ is the perturbative parameter and $g_{\mu \nu} \rightarrow \eta_{\mu \nu}$ for $r \rightarrow \infty$. We can see the Newtonian limit represented by $\phi$ at order $\epsilon^2$ in the components of $g_{\mu \nu}$. In the Post-Newtonian limit, we have ten additional functions to represent ten degrees of freedom on the metric. In the same way, $\tilde{g}_{\mu \nu}$ will be:

\begin{eqnarray}
\label{gt PostNewton} \tilde{g}_{\mu \nu}dx^{\mu}dx^{\nu} &=&  - 2\left(\tilde{\phi}\epsilon^2+\left(2\phi\tilde{\phi} + \tilde{\psi}\right)\epsilon^4\right) \left(\frac{c dt}{\epsilon}\right)^2 - 2\left(\tilde{\phi}\epsilon^2 + \tilde{\psi}\epsilon^4\right)\left(dx^2+dy^2+dz^2\right) \nonumber \\
&& + 2\epsilon^3\left(\tilde{\chi}_1dx + \tilde{\chi}_2dy + \tilde{\chi}_3dz\right) \left(\frac{c dt}{\epsilon}\right) + \epsilon^4\left(\tilde{\xi}_{11}dx^2 + \tilde{\xi}_{22}dy^2 + \tilde{\xi}_{33}dz^2\right. \nonumber \\
&& \left. + 2\tilde{\xi}_{12} dx dy + 2\tilde{\xi}_{13} dx dz + 2\tilde{\xi}_{23} dy dz\right),
\end{eqnarray}

where we used $\tilde{g}_{\mu \nu} \rightarrow 0$ for $r \rightarrow \infty$. All functions in (\ref{g PostNewton}) and (\ref{gt PostNewton}) depend of $(t,x,y,z)$, but $\frac{1}{c}\frac{\partial}{\partial t} \sim \epsilon$. For this reason, we use $c t \rightarrow \frac{c t}{\epsilon}$ to obtain the equations. (\ref{g PostNewton}) and (\ref{gt PostNewton}) are the more general expression for a covariant tensor of rank two, so we need to fix a gauge. One particularly convenient gauge is given by the extended harmonic coordinate conditions. It must be extended because we need to consider the $\tilde{g}_{\mu \nu}$'s components. Then, the gauge fixing is given by (For more details, see \cite{Paper 1}):

\begin{eqnarray}
\label{Harmonic gauge}
\Gamma^{\mu} &\equiv& g^{\alpha \beta}\Gamma^{\mu}_{\alpha \beta} = 0 \\
\label{Harmonic gauge tilde}
\tilde{\delta}\left(\Gamma^{\mu}\right) &\equiv& g^{\alpha \beta}\tilde{\delta}\left(\Gamma^{\mu}_{\alpha \beta}\right) - \tilde{g}^{\alpha \beta}\Gamma^{\mu}_{\alpha \beta} = 0,
\end{eqnarray}

where $\tilde{\delta}\left(\Gamma^{\mu}_{\alpha \beta}\right) = \frac{1}{2}g^{\mu \lambda}\left(D_{\beta}\tilde{g}_{\lambda \alpha}+D_{\alpha}\tilde{g}_{\beta \lambda}-D_{\lambda}\tilde{g}_{\alpha \beta}\right)$. Applying (\ref{Harmonic gauge}) and (\ref{Harmonic gauge tilde}) to (\ref{g PostNewton}) and (\ref{gt PostNewton}), we obtain:

\begin{eqnarray}
\label{gauge PN 1}
4 \dot{\phi} + \partial_i \chi_i &=& 0 \\
\label{gauge PN 2}
2 \phi \partial_i \phi - \dot{\chi}_i - \frac{1}{2}\partial_i \xi_{jj} + \partial_j \xi_{ij} &=& 0 \\
\label{gauge PN 3}
4 \dot{\tilde{\phi}} + \partial_i \tilde{\chi}_i &=& 0 \\
\label{gauge PN 4}
2 \phi \partial_i \tilde{\phi} + 2 \tilde{\phi} \partial_i \phi - \dot{\tilde{\chi}}_i - \frac{1}{2}\partial_i \tilde{\xi}_{jj} + \partial_j \tilde{\xi}_{ij} &=& 0.
\end{eqnarray}

On the other side, in \cite{Paper 1} is proved that the energy-momentum tensors for a perfect fluid are given by:

\begin{eqnarray}
\label{T PF}
T_{\mu \nu} &=& p(\rho)g_{\mu \nu} + \left(\rho + p(\rho)\right)U_{\mu}U_{\nu} \\[5pt]
\label{T tilde PF}
\tilde{T}_{\mu \nu}
&=& p(\rho)\tilde{g}_{\mu \nu} + \frac{\partial p}{\partial \rho}(\rho)\tilde{\rho}g_{\mu \nu} + \left(\tilde{\rho} + \frac{\partial p}{\partial \rho}(\rho)\tilde{\rho}\right)U_{\mu}U_{\nu} \nonumber \\
&& + \left(\rho + p(\rho)\right)\left(\frac{1}{2}\left(U_{\nu}U^{\alpha}\tilde{g}_{\mu \alpha} + U_{\mu}U^{\alpha}\tilde{g}_{\nu \alpha}\right)+U^T_{\mu}U_{\nu}+U_{\mu}U^T_{\nu}\right),
\end{eqnarray}

where $U^{\mu}U_{\mu} = -1$ and $U^{\mu}U^T_{\mu} = 0$. Now, in the Non-Relativistic limit, we have:

\begin{eqnarray}
\rho &=& \rho^{(0)} + \epsilon^2 \rho^{(2)} \\[5pt]
\tilde{\rho} &=& \tilde{\rho}^{(0)} + \epsilon^2 \tilde{\rho}^{(2)} \\[5pt]
p(\rho) &=& \epsilon^2 p^{(2)}(\rho) \\[5pt]
U_{\mu} &=& \left(c\left(\frac{1+\epsilon^2\left(\phi + \frac{1}{2}U^{(1)}_kU^{(1)}_k\right)}{\epsilon}\right),\epsilon U^{(1)}_i\right) \\[5pt]
U^T_{\mu} &=& \left(c\epsilon U^{T(1)}_kU^{(1)}_k,\epsilon U^{T(1)}_i\right).
\end{eqnarray}

With all these, the equations of motion (\ref{Einst Eq}) and (\ref{tilde Eq}) are given by:

\begin{eqnarray}
\label{Eq PN 01}
\partial^2 \phi &=& \frac{\kappa}{2} \rho^{(0)} \\[5pt]
\label{Eq PN 02}
\partial^2 \chi_i &=& - 2\kappa U_i^{(1)}\rho^{(0)} \\[5pt]
\label{Eq PN 03}
\partial^2 \psi &=& \frac{\kappa}{2} \left(2\left(U_k^{(1)}U_k^{(1)}-\phi\right)\rho^{(0)} + \rho^{(2)} + 3p^{(2)}(\rho)\right) + \ddot{\phi} \\[5pt]
\label{Eq PN 04}
\partial^2 \xi_{ij} &=& - 2\kappa U_i^{(1)}U_j^{(1)}\rho^{(0)} - 4(\partial_i \phi)(\partial_j \phi) + 2\kappa\left(\left(U_k^{(1)}U_k^{(1)}+\phi\right)\rho^{(0)}+2p^{(2)}(\rho)\right)\delta_{ij} \nonumber \\
&& + 4(\partial_k \phi)(\partial_k \phi)\delta_{ij} \\[5pt]
\label{Eq PN 05}
\partial^2 \tilde{\phi} &=& \frac{\kappa}{2} \tilde{\rho}^{(0)} \\[5pt]
\label{Eq PN 06}
\partial^2 \tilde{\chi}_i &=& - 2\kappa \left(U_i^{T(1)}\rho^{(0)} + U_i^{(1)}\tilde{\rho}^{(0)}\right) \\[5pt]
\label{Eq PN 07}
\partial^2 \tilde{\psi} &=& \kappa \left(\left(2U_k^{(1)}U_k^{T(1)}-\tilde{\phi}\right)\rho^{(0)} + \left(U_k^{(1)}U_k^{(1)}-\phi+\frac{3}{2}p'^{(2)}(\rho)\right)\tilde{\rho}^{(0)} + \frac{\tilde{\rho}^{(2)}}{2}\right) + \ddot{\tilde{\phi}} \\[5pt]
\label{Eq PN 08}
\partial^2 \tilde{\xi}_{ij} &=& - 2\kappa\left(\left(U_i^{T(1)}U_j^{(1)}+U_i^{(1)}U_j^{T(1)}\right)\rho^{(0)} + U_i^{(1)}U_j^{(1)}\tilde{\rho}^{(0)}\right) - 4(\partial_i \tilde{\phi})(\partial_j \phi) - 4(\partial_i \phi)(\partial_j \tilde{\phi}) \nonumber \\&& + 2\kappa\left(\left(2U_k^{(1)}U_k^{T(1)}+\tilde{\phi}\right)\rho^{(0)}+\left(U_k^{(1)}U_k^{(1)}+\phi+2p'^{(2)}(\rho)\right)\tilde{\rho}^{(0)}\right)\delta_{ij} \nonumber \\&& + 8(\partial_k \phi)(\partial_k \tilde{\phi})\delta_{ij},
\end{eqnarray}

where $p'^{(2)}(\rho) = \frac{\partial p^{(2)}}{\partial \rho}(\rho)$, $\partial^2 = \partial_i \partial_i$, $\partial_i = \frac{\partial}{\partial x^i}$ and $\kappa \sim O(\epsilon^2)$. We can see that the equations (\ref{Eq PN 01}) and (\ref{Eq PN 05}) correspond to the Newtonian limit. To complete the system, we have the equations (\ref{Conserv T}) and (\ref{Conserv tilde T}), but they are satisfied when we consider (\ref{gauge PN 1}-\ref{gauge PN 4}). However, they are useful because we can write them in terms of $\rho^{(0)}$, $\rho^{(2)}$, $\tilde{\rho}^{(0)}$, $\tilde{\rho}^{(2)}$ and $p^{(2)}$. In spherical symmetry with $U^{(1)}_i = U^{T(1)}_i = 0$, it is possible to prove that the conservation equations can be reduced to:

\begin{eqnarray}
\label{Cons EQ PN 1}
\frac{\partial p^{(2)}}{\partial r}(r) + \rho^{(0)}(r)\left(\frac{\partial \phi}{\partial r}(r)\right) &=& 0 \\
\label{Cons EQ PN 2}
\frac{\partial}{\partial r}\left(\left(1+2\epsilon^2\phi(r)\right)^{-1}\left(\tilde{\phi}(r) - \tilde{\rho}^{(0)}(r)\frac{\left(\frac{\partial \phi}{\partial r}(r)\right)}{\left(\frac{\partial \rho^{(0)}}{\partial r}(r)\right)}\right)\right) &=& 0.
\end{eqnarray}

From (\ref{Cons EQ PN 1}) we can obtain $p^{(2)}(r)$ and (\ref{Cons EQ PN 2}) say us that:

\begin{eqnarray}
\label{tilde rho0}
\tilde{\rho}^{(0)}(r) = \frac{\left(\frac{\partial \rho^{(0)}}{\partial r}(r)\right)}{\left(\frac{\partial \phi}{\partial r}(r)\right)}\left(\tilde{\phi}(r) - C\left(1+2\epsilon^2\phi(r)\right)\right),
\end{eqnarray}

where $C$ is an integration constant. We preserve $2\epsilon^2\phi(r)$ in the last term of the right side because the order of $C$ is unknown. Equations (\ref{Cons EQ PN 1}) and (\ref{Cons EQ PN 2}) are obtained in a Post-Newtonian level. This means that we need to study the system at this level to obtain the relation (\ref{tilde rho0}) and complete the Newtonian limit. Finally, the equations (\ref{Eq PN 01}) and (\ref{Eq PN 05}) for spherical symmetry are reduced to:

\begin{eqnarray}
\label{Newton Eq 1b}
\frac{1}{r^2}\frac{\partial}{\partial r}\left(r^2\left(\frac{\partial \phi(r)}{\partial r}\right)\right) &=& \frac{\kappa}{2} \rho^{(0)}(r) \\
\label{Newton Eq 2b}
\frac{1}{r^2}\frac{\partial}{\partial r}\left(r^2\left(\frac{\partial \tilde{\phi}(r)}{\partial r}\right)\right) &=& \frac{\kappa}{2} \frac{\left(\frac{\partial \rho^{(0)}}{\partial r}(r)\right)}{\left(\frac{\partial \phi}{\partial r}(r)\right)}\left(\tilde{\phi}(r) - C\left(1+2\epsilon^2\phi(r)\right)\right).
\end{eqnarray}

Then, to obtain $\phi(r)$ and $\tilde{\phi}(r)$, we just need $\rho^{(0)}(r)$, completing the Newtonian limit. Now, we can ask ourselves if it is possible to explain Dark Matter with this result. For this, we will need to study the trajectory of a free particle.\\

\subsection{\label{SubSec: trajectory of a Particle}Trajectory of a Particle:}

From \textbf{Appendix B}, we know that the acceleration is given by (\ref{geodesics m}). In the Post-Newtonian limit, we obtain that \cite{Weinberg grav}:

\begin{eqnarray}
\label{acceleration PN 0}
\frac{1}{c^2}\frac{d^2 \vec{x}}{d t^2} &=& - \epsilon^2 \nabla \left(\phi_N+ \left(2\phi_N^2 + \psi_N\right)\epsilon^2\right) \nonumber \\
&& + \epsilon^4 \left(3\vec{v}\dot{\phi}_N + 4\vec{v}\left(\vec{v} \cdot \nabla \phi_N\right) - v^2 \nabla \phi_N - \dot{\vec{\chi}}_N + \left(\vec{v} \times \nabla \times \vec{\chi}_N\right)\right) \nonumber \\
&& + \frac{\epsilon^4}{2} \nabla \tilde{\phi}^2 + O\left(\epsilon^6\right),
\end{eqnarray}

where $\vec{v} = \frac{d \vec{x}}{d t}$, $\phi_N = \phi + \tilde{\phi}$ and analogous expressions for the others fields. From (\ref{acceleration PN 0}), we can deduce a couple of things. Firstly, in the Newtonian limit we have:

\begin{eqnarray}
\label{acceleration PN}
\frac{1}{c^2}\frac{d^2 \vec{x}}{d t^2} &=& - \epsilon^2 \nabla \phi_N,
\end{eqnarray}

so $\phi_N$ is the effective Newtonian potential. Secondly, the acceleration is similar to the usual case if we replace $\phi \rightarrow \phi_N$, with the exception of the last term in (\ref{acceleration PN 0}). If we analyze the case with spherical symmetry far away from matter, we can see from (\ref{Newton Eq 2b}) that $\tilde{\phi}^2 \sim r^{-2}$. This means that this term is $\sim - r^{-3}$, therefore it is an attractive contribution and can be considered as a contribution to Dark Matter. In this paper, we will focus in the Newtonian approximation given by (\ref{acceleration PN}), which is the dominant term.\\

We have said that $\phi_N$ is the effective potential in the Newtonian limit. This means that the effective density is $\rho_{eff} = \rho^{(0)} + \tilde{\rho}^{(0)}$. In spherical symmetry, that is:

\begin{eqnarray}
\label{rho eff}
\rho_{eff}(r) = \rho^{(0)}(r) + \frac{\left(\frac{\partial \rho^{(0)}}{\partial r}(r)\right)}{\left(\frac{\partial \phi}{\partial r}(r)\right)}\left(\tilde{\phi}(r) - C\left(1+2\epsilon^2\phi(r)\right)\right).
\end{eqnarray}

Therefore, the second term in (\ref{rho eff}) is an additional mass and it could be identified as Dark Matter. To verify this, we need to use some density profile for any galaxy and then obtain the effective density. Next, we will study (\ref{rho eff}) and the equations of motion of $\phi$ and $\tilde{\phi}$ to see if $\tilde{\delta}$ Matter can explain the Dark Matter effect.\\[30pt]

\section{\label{Sec: Density Profiles}Density Profiles:}

To study the $\tilde{\delta}$ Matter effects we must analyze the equations (\ref{tilde rho0}-\ref{Newton Eq 2b}). To explain Dark Matter, $\tilde{\delta}$ Matter must be negligible in the solar system scale, but important in galactic scale. We will study these equations with some density profile. In the first place, we will see a spherically homogeneous density like a first approximation for a planetary or stellar distribution. Then, we will study the exponential profile and finally we will use the Einasto and Navarro-Frenk-White (NFW) profiles to describe galaxies distributions.\\

We define a normalized radius $x$ such that $r = Rx$, then our equations are:

\begin{eqnarray}
\label{eq newton 1}
\frac{1}{x^2}\frac{\partial}{\partial x}\left(x^2\left(\frac{\partial \phi(x)}{\partial x}\right)\right) &=& \frac{\kappa R^2}{2} \rho(x) \\
\label{eq newton 2}
\frac{1}{x^2}\frac{\partial}{\partial x}\left(x^2\left(\frac{\partial \tilde{\phi}(x)}{\partial x}\right)\right) &=& \frac{\kappa R^2}{2}\frac{\left(\frac{\partial \rho}{\partial x}(x)\right)}{\left(\frac{\partial \phi}{\partial x}(x)\right)}\left(\tilde{\phi}(x) - C\left(1+2\epsilon^2\phi(x)\right)\right),
\end{eqnarray}

where $R$ is a convenient radius. With these equations, we can define the ordinary mass and tilde mass. That is:

\begin{eqnarray}
\label{m ord}
m(x) &=& 4\pi R^3 \int_0^{\infty} dx x^2\rho(x) \nonumber \\
&=& \frac{8\pi R}{\kappa} x^2\left(\frac{\partial \phi(x)}{\partial x}\right) \\
\label{m tild}
\tilde{m}(x) &=& 4\pi R^3 \int_0^{\infty} dx x^2\tilde{\rho}(x) \nonumber \\
&=& \frac{8\pi R}{\kappa} x^2\left(\frac{\partial \tilde{\phi}(x)}{\partial x}\right),
\end{eqnarray}

such that the effective mass is $M(x) = m(x) + \tilde{m}(x)$. Finally, from (\ref{acceleration PN}), we have that the rotation velocity is:

\begin{eqnarray}
\label{vel Galax}
\left(\frac{v_{rot}(x)}{c}\right)^2 &=& \epsilon^2 x\frac{\partial}{\partial x}\left(\phi(x) + \tilde{\phi}(x)\right) \nonumber \\
&=& \frac{\epsilon^2\kappa M(x)}{8\pi R x},
\end{eqnarray}

where we used that $\frac{d^2 \vec{x}}{d t^2} = - \frac{v_{rot}^2(r)}{r}\hat{\theta}$ in this case.\\

\subsection{\label{SubSec:Spherically Homogeneous Profile}Spherically Homogeneous Profile:}

We can think, in a first approximation, that planets or stars are spheres with a constant density. That is $\rho(x) = \rho_0\Theta(1-x)$. So, (\ref{eq newton 1}) and (\ref{eq newton 2}) are:

\begin{eqnarray}
\frac{1}{x^2}\frac{\partial}{\partial x}\left(x^2\left(\frac{\partial \phi(x)}{\partial x}\right)\right) &=& \frac{\kappa R^2 \rho_0}{2} \Theta(1-x) \nonumber \\
\frac{1}{x^2}\frac{\partial}{\partial x}\left(x^2\left(\frac{\partial \tilde{\phi}(x)}{\partial x}\right)\right) &=& - \frac{\kappa R^2 \rho_0}{2} \frac{\delta(1-x)}{\left(\frac{\partial \phi}{\partial x}(x)\right)}\left(\tilde{\phi}(x) - C\left(1+2\epsilon^2\phi(x)\right)\right) \nonumber \\
\tilde{\rho}(x) &=& - \rho_0 \frac{\delta(1-x)}{\left(\frac{\partial \phi}{\partial x}(x)\right)}\left(\tilde{\phi}(x) - C\left(1+2\epsilon^2\phi(x)\right)\right), \nonumber
\end{eqnarray}

where $R$ is the radius of the sphere. From the first equation, we can obtain $\phi(x)$. Using the boundary conditions $\phi(\infty) \rightarrow 0$, $\phi(0) \rightarrow \textit{"finity value"}$ and imposing that $\phi(x)$ and $\phi'(x)$ must be continuous for all $x$, we obtain:

\begin{eqnarray}
\label{Sol 01 phi}
\phi(x) &=& \left\{
\begin{array}{cc}
  \frac{\kappa R^2 \rho_0}{12}\left(x^2-3\right) & x \leq 1 \\
  - \frac{\kappa R^2 \rho_0}{6x} & x > 1
\end{array}\right..
\end{eqnarray}

On the other side, from the second equation, we obtain $\tilde{\phi}(x)$, but we can not impose a continuous $\tilde{\phi}'(x)$. In its stead the equation of motion of $\tilde{\phi}(x)$ say us:

\begin{eqnarray}
\left.x^2\left(\frac{\partial \tilde{\phi}(x)}{\partial x}\right)\right|_{int}^{ext} &=& - \frac{\kappa R^2 \rho_0}{2\phi'(1)}\left(\tilde{\phi}(1) - C\left(1+2\epsilon^2\phi(1)\right)\right), \nonumber
\end{eqnarray}

where the other condition on $\tilde{\phi}(x)$ are the same. That is $\tilde{\phi}(\infty) \rightarrow 0$, $\tilde{\phi}(0) \rightarrow \textit{"finity value"}$ and $\tilde{\phi}(x)$ is continuous for all $x$. Then, the solution is:

\begin{eqnarray}
\label{Sol 01 tphi}
\tilde{\phi}(x) &=& \left\{
\begin{array}{cc}
  \frac{3}{2}C\left(1-\frac{\epsilon^2 \kappa R^2 \rho_0}{3}\right) & x \leq 1 \\
  \frac{3}{2x}C\left(1-\frac{\epsilon^2 \kappa R^2 \rho_0}{3}\right) & x > 1
\end{array}\right..
\end{eqnarray}

Besides, the $\tilde{\delta}$ Matter density is:

\begin{eqnarray}
\label{Sol 01 trho}
\tilde{\rho}(x) &=& - \frac{3C}{\kappa R^2}\left(1-\frac{\epsilon^2 \kappa R^2 \rho_0}{3}\right)\delta(x-1) \nonumber \\
&\approx& - \frac{3C}{\kappa R^2}\delta(x-1).
\end{eqnarray}

Besides, the acceleration is given by (\ref{acceleration PN}), then:

\begin{eqnarray}
\vec{a} = - \epsilon^2 \frac{c^2}{R} \frac{\partial}{\partial x}\left(\phi(x) + \tilde{\phi}(x)\right)\hat{r}
\end{eqnarray}

Naturally, we expect a continuous $\vec{a}$, but we saw that $\tilde{\phi}'(x)$ is not. This means that we have to accept that $\tilde{\delta}$ Gravity produce an additional force on the surface of the sphere, or $C = 0$. In the last case, all $\tilde{\delta}$ components disappear, so $\tilde{\delta}$ Gravity is the same as GR. This is a really important condition, so this result suggests that $\tilde{\delta}$ Gravity is negligible with this density profile.\\

In conclusion, we obtain that an spherically homogeneous distribution of ordinary matter, where the density is constant by $x<1$ and then go to $0$ instantly, does not have $\tilde{\delta}$ Matter. This case can be an acceptable representation of planets and stars, where $\tilde{\delta}$ Matter does not produce important effects. However, $\tilde{\delta}$ Matter must be important when the distribution of ordinary matter is variable, which means that in a galaxy we can obtain important effects where $\tilde{\delta}$ Matter has a behavior as Dark Matter. Now we will study standard distributions used to represent the density in a galaxy and we will see new effects produced by $\tilde{\delta}$ Gravity.\\

\subsection{\label{SubSec:Exponential Profile}Exponential Profile:}

\begin{figure}
\centering
\includegraphics[scale=0.6]{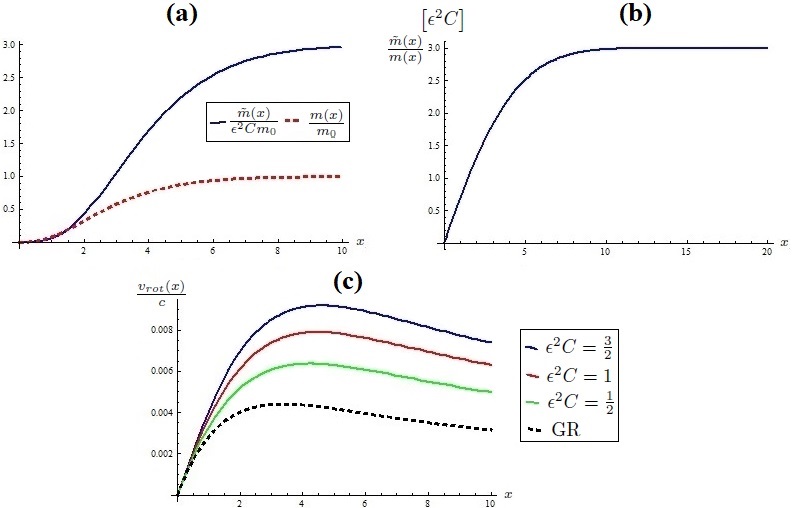}
\caption{{\scriptsize Exponential Profile Calculation. \textbf{(a)} Ordinary and $\tilde{\delta}$ Matter vs normalize radius, with $m_0 = 8\pi R^3 \rho_0$. From the center, more $\tilde{\delta}$ Matter is accumulated, but the behavior of both terms become similar when we distance from the center. \textbf{(b)} $\frac{\tilde{m}(x)}{m(x)}$ (in units of $\epsilon^2C$) vs normalize radius. We verify the last conclusion. In the center the relation is almost linear, $\frac{\tilde{m}(x)}{m(x)} \sim x$, and in the end of the galaxy it is like a constant, $\frac{\tilde{m}(x)}{m(x)} \rightarrow 3\epsilon^2C$. \textbf{(c)} Rotation velocity vs normalize radius for different values of $\epsilon^2C$. The Black-Dashed line corresponds to GR case, so the $\epsilon^2C$ value indicates the contribution of $\tilde{\delta}$ Matter. The similar behavior of both elements and the additional mass from $\tilde{\delta}$ Matter produce an amplifying effect in the rotation velocity. In these calculations we have used $\kappa \rho_0 R^2 = 10^{-4}$}.}
\label{Fig: Exp Sol}
\end{figure}

We will study this profile to develop an acceptable first approximation to a galaxy distribution. That is $\rho(x) = \rho_0 e^{-x}$. In this case, (\ref{eq newton 1}) and (\ref{eq newton 2}) are:

\begin{eqnarray}
\frac{1}{x^2}\frac{\partial}{\partial x}\left(x^2\left(\frac{\partial \phi(x)}{\partial x}\right)\right) &=& \frac{\kappa R^2 \rho_0 e^{-x}}{2} \nonumber \\
\frac{1}{x^2}\frac{\partial}{\partial x}\left(x^2\left(\frac{\partial \tilde{\phi}(x)}{\partial x}\right)\right) &=& - \frac{\kappa R^2 \rho_0 e^{-x}}{2\left(\frac{\partial \phi}{\partial x}(x)\right)}\left(\tilde{\phi}(x) - C\left(1+2\epsilon^2\phi(x)\right)\right). \nonumber
\end{eqnarray}

From the first equation, we can solve $\phi(x)$ analytically. Using the boundary conditions $\phi(\infty) \rightarrow 0$ and $\phi(0) \rightarrow \textit{"finity value"}$, we obtain:

\begin{eqnarray}
\label{sol phi exp}
\phi(x) &=& \frac{\kappa R^2 \rho_0 \left(\left(x+2\right)e^{-x}-2\right)}{2x}.
\end{eqnarray}

On the other side, the equation of $\tilde{\phi}(x)$ is too complicated. That is:

\begin{eqnarray}
\label{sol tphi exp}
\frac{\partial}{\partial x}\left(x^2\left(\frac{\partial \tilde{\phi}(x)}{\partial x}\right)\right) &=& - \frac{x^3e^{-x}\left(x\tilde{\phi}(x) - C\left(x - \epsilon^2 \kappa R^2 \rho_0 \left(2-\left(x+2\right)e^{-x}\right)\right)\right)}{2-\left(x^2+2x+2\right)e^{-x}}.
\end{eqnarray}

Now we can solve this equation numerically and find $m(x)$ and $\tilde{m}(x)$. First we must analyze the initial conditions. We can do it studying the behavior of $m(x)$ and $\tilde{m}(x)$, given by (\ref{m ord}) and (\ref{m tild}), for $x \ll 1$ because they are related to $x^2\left(\frac{\partial \phi(x)}{\partial x}\right)$ and $x^2\left(\frac{\partial \tilde{\phi}(x)}{\partial x}\right)$ respectively. From (\ref{sol phi exp}), we have that $\phi(x) \approx \frac{\kappa R^2 \rho_0}{2}$ and $m(x) \approx \frac{4\pi R^3 \rho_0}{3}x^3$ for small $x$. On the other side, from equation (\ref{sol tphi exp}), we need $\tilde{\phi}(x) \approx C\left(1-\epsilon^2 \kappa R^2 \rho_0\right)$ and $\tilde{m}(x) \approx \epsilon^2C\pi R^3 \rho_0 x^4$. So, the total mass is completely dominated by ordinary components in the center.\\

Previously, we said that the $C$ order is unknown, but the initial conditions say us that $\frac{\tilde{m}(x)}{m(x)} \approx \frac{3\epsilon^2C}{4}x$. This means that, the $\tilde{\delta}$ Matter is increased from $x<<1$ too slowly, unless $\epsilon^2C \sim O(1)$. We can see this in \textbf{Figure \ref{Fig: Exp Sol}a}, where we present to $m(x)$ and $\tilde{m}(x)$. Additionally, we represent the relation between $m(x)$ and $\tilde{m}(x)$ in \textbf{Figure \ref{Fig: Exp Sol}b}. These plots say us that ordinary matter is dominant in the center, but $\tilde{\delta}$ Matter is accumulated when we get away from the center and the behavior of both kinds of matters become similar when $x$ increase. In fact, the relation between them is practically constant in the edge of the galaxy, $\frac{\tilde{m}(x)}{m(x)} \rightarrow 3\epsilon^2C$. In conclusion, ordinary and $\tilde{\delta}$ Matter evolve in a similar way, but $\tilde{\delta}$ Matter is concentrated outside of the galactic nucleus.\\

In \textbf{Figure \ref{Fig: Exp Sol}c}, we show the rotation velocity in $\tilde{\delta}$ Gravity with $\epsilon^2C = \left\{\frac{1}{2}, 1, \frac{3}{2}\right\}$ and GR. Clearly, the similar behavior of both elements and the additional mass from $\tilde{\delta}$ Matter produce an amplifying effect in the rotation velocity. This means that a minimal quantity of ordinary matter could explain the rotation velocity in a galaxy, because of the additional contribution produced by $\tilde{\delta}$ Matter.\\

\subsection{\label{SubSec:Einasto Profile}Einasto Profile:}

The Einasto profile is a spherically symmetric distribution used to describe many types of real system, like galaxies and Dark Matter halos (see for instance \cite{Einasto 1}-\cite{Einasto 3}). It is represented by a logarithmic power-law:

\begin{eqnarray}
\frac{d \ln(\rho(r))}{d \ln r} \propto - r^{\alpha}, \nonumber
\end{eqnarray}

with $\alpha > 0$, then $\rho(x) = \rho_0 e^{-x^{\alpha}}$. So, it is a most general case of the exponential profile and many simulations of galaxies have been done using this profile, where they obtained values of $\alpha$ given by $0.1 \leq \alpha \leq 1$ \cite{Einasto 3}. Evaluating in (\ref{eq newton 1}) and (\ref{eq newton 2}), we have:

\begin{eqnarray}
\label{sol phi einasto}
\frac{1}{x^2}\frac{\partial}{\partial x}\left(x^2\left(\frac{\partial \phi(x)}{\partial x}\right)\right) &=& \frac{\kappa R^2 \rho_0 e^{-x^{\alpha}}}{2} \\
\label{sol tphi einasto}
\frac{1}{x^2}\frac{\partial}{\partial x}\left(x^2\left(\frac{\partial \tilde{\phi}(x)}{\partial x}\right)\right) &=& - \frac{\kappa R^2 \rho_0 \alpha x^{\alpha-1} e^{-x^{\alpha}}}{2\left(\frac{\partial \phi}{\partial x}(x)\right)}\left(\tilde{\phi}(x) - C\left(1+2\epsilon^2\phi(x)\right)\right).
\end{eqnarray}

\begin{figure}
\centering
\includegraphics[scale=0.55]{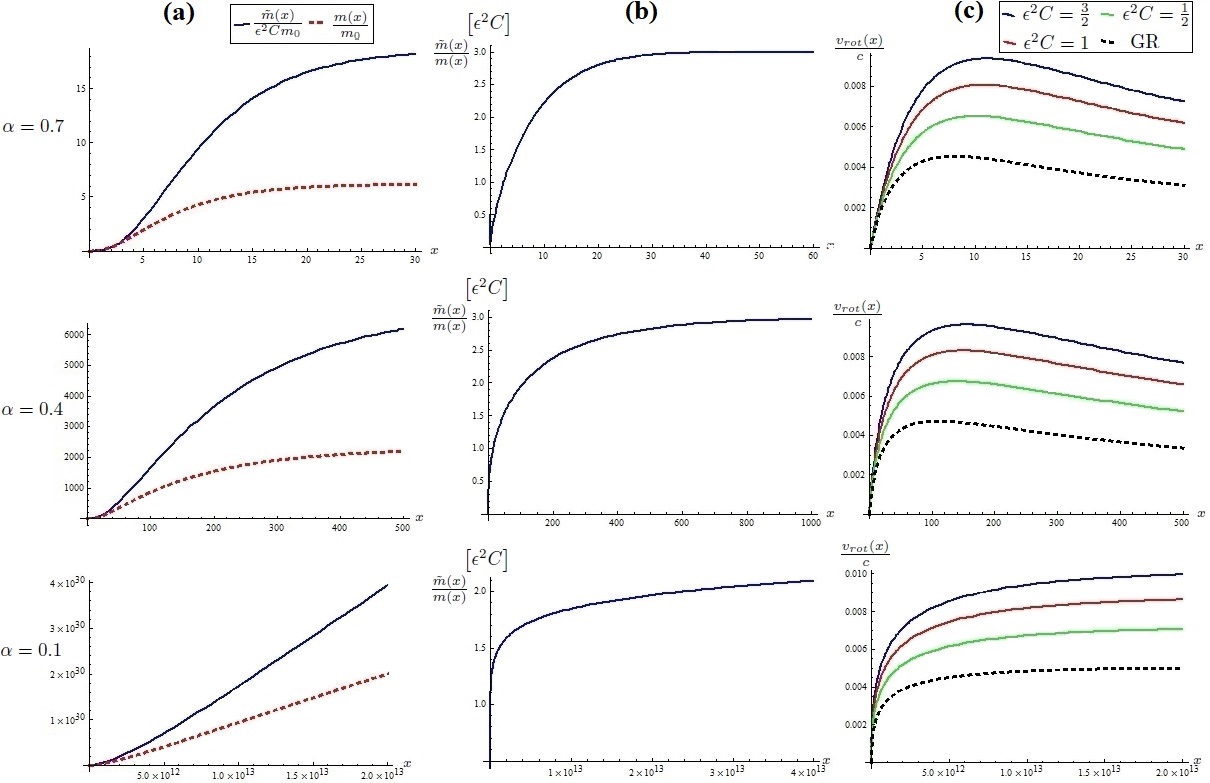}
\caption{{\scriptsize Einasto Profile Calculation for $\alpha = \{0.7,0.4,0.1\}$. \textbf{(a)} Ordinary and $\tilde{\delta}$ Matter vs normalize radius, with $m_0 = 8\pi R^3 \rho_0$. From the center, more $\tilde{\delta}$ Matter is accumulated, just like the exponential case. \textbf{(b)} $\frac{\tilde{m}(x)}{m(x)}$ (in units of $\epsilon^2C$) vs normalize radius. In fact, we have that $\tilde{m}(x)$ increases faster than $m(x)$ and it is faster when $\alpha$ is smaller. On the other side, in the end of the galaxy it is like a constant too, with $\frac{\tilde{m}(x)}{m(x)} \rightarrow 3\epsilon^2C$, but for smaller $\alpha$ the behavior is more similar to a logarithmic function. \textbf{(c)} Rotation velocity vs normalize radius for different values of $\epsilon^2C$. The Black-Dashed line corresponds to GR case, so the $\epsilon^2C$ value indicates the contribution of $\tilde{\delta}$ Matter. In this case we have an amplifying effect in the rotation velocity too, such that if $\epsilon^2C$ is bigger, we have higher velocities. In these calculations we have used $\kappa \rho_0 R^2 = 4\left(\frac{\alpha}{2}\right)^{\frac{2}{\alpha}}e^{\frac{2\left(1-\alpha\right)}{\alpha}} \times 10^{-4}$.}}
\label{Fig: Einasto Sol}
\end{figure}

As in the exponential profile, we can solve (\ref{sol phi einasto}) and (\ref{sol tphi einasto}) to find $m(x)$ and $\tilde{m}(x)$\footnote{The only constant that we can not fix is $\phi_0 = \phi(0)$. Fortunately, this constant is irrelevant to find $m(x)$. This is true for NFW profile too.}. So, using (\ref{m ord}) and (\ref{m tild}), we can see that the appropriate initial conditions are given by $m(x) \approx \frac{4\pi R^3 \rho_0}{3}x^3$ and $\tilde{m}(x) \approx \frac{4\pi R^3 \rho_0 \epsilon^2 C\alpha}{\left(3+\alpha\right)}x^{\alpha+3}$. Clearly, we can verify that this result is reduced to the exponential case with $\alpha = 1$ and we need $\epsilon^2C \sim O(1)$ to obtain enough $\tilde{\delta}$ Matter too. $m(x)$ and $\tilde{m}(x)$ are represented in \textbf{Figure \ref{Fig: Einasto Sol}a} for $\alpha = \{0.7,0.4,0.1\}$, and we represent the relation between $m(x)$ and $\tilde{m}(x)$ in \textbf{Figure \ref{Fig: Einasto Sol}b}. As in the exponential case, more $\tilde{\delta}$ Matter is accumulated far from the center. Actually, we have that $\tilde{m}(x)$ increases faster than $m(x)$, specially when $\alpha$ is smaller. However, far from the center, $\frac{\tilde{m}(x)}{m(x)} \rightarrow \textit{constant} \leq 3\epsilon^2C$ when $\alpha$ is close to $1$. For smaller $\alpha$'s, it is more like a logarithmic behavior. Finally, in \textbf{Figure \ref{Fig: Einasto Sol}c}, we present the rotation velocity for $\epsilon^2C = \left\{\frac{1}{2}, 1, \frac{3}{2}\right\}$ and GR. Just like we expected, the rotation velocity is amplified by the $\tilde{\delta}$ Matter effect, in such a way that if $C$ is bigger, we have higher velocities.\\

This result give us the same conclusion that exponential profile. The ordinary matter leads over $\tilde{\delta}$ Matter in the center, but rapidly the second one increase until it is completely dominant. So, $\tilde{\delta}$ Matter is also concentrated outside of the galactic nucleus in this case, but additionally we obtained a logarithmic contribution from $\tilde{\delta}$ Matter, producing an additional Dark Matter behavior to small values of $\alpha$. We note this in \textbf{Figure \ref{Fig: Einasto Sol}c} for $\alpha = 0.1$.\\

\subsection{\label{SubSec:Navarro-Frenk-White Profile}Navarro-Frenk-White Profile:}

The Navarro-Frenk-White (NFW) profile is another kind of distribution of the of Dark Matter halo (see for instance \cite{NFW 1}-\cite{NFW 3}), given by $\rho(x) = \frac{\rho_0}{x^{\gamma}(x+1)^{3-\gamma}}$. In pure Dark Matter simulations $\gamma = 1$ is usually used, however baryonic matter effects are expected, producing $1 \leq \gamma \leq 1.4$ \cite{NFW 3}. Now, in this case, (\ref{eq newton 1}) and (\ref{eq newton 2}) are:

\begin{eqnarray}
\label{sol phi NWF}
\frac{1}{x^2}\frac{\partial}{\partial x}\left(x^2\left(\frac{\partial \phi(x)}{\partial x}\right)\right) &=& \frac{\kappa R^2 \rho_0}{2x^{\gamma}\left(x+1\right)^{3-\gamma}} \\
\label{sol tphi NFW}
\frac{1}{x^2}\frac{\partial}{\partial x}\left(x^2\left(\frac{\partial \tilde{\phi}(x)}{\partial x}\right)\right) &=& - \frac{\kappa R^2 \rho_0 \left(\gamma+3x\right)}{2x^{\gamma+1}\left(x+1\right)^{4-\gamma}\left(\frac{\partial \phi}{\partial x}(x)\right)}\left(\tilde{\phi}(x) - C\left(1+2\epsilon^2\phi(x)\right)\right).
\end{eqnarray}

\begin{figure}
\begin{center}
\includegraphics[scale=0.55]{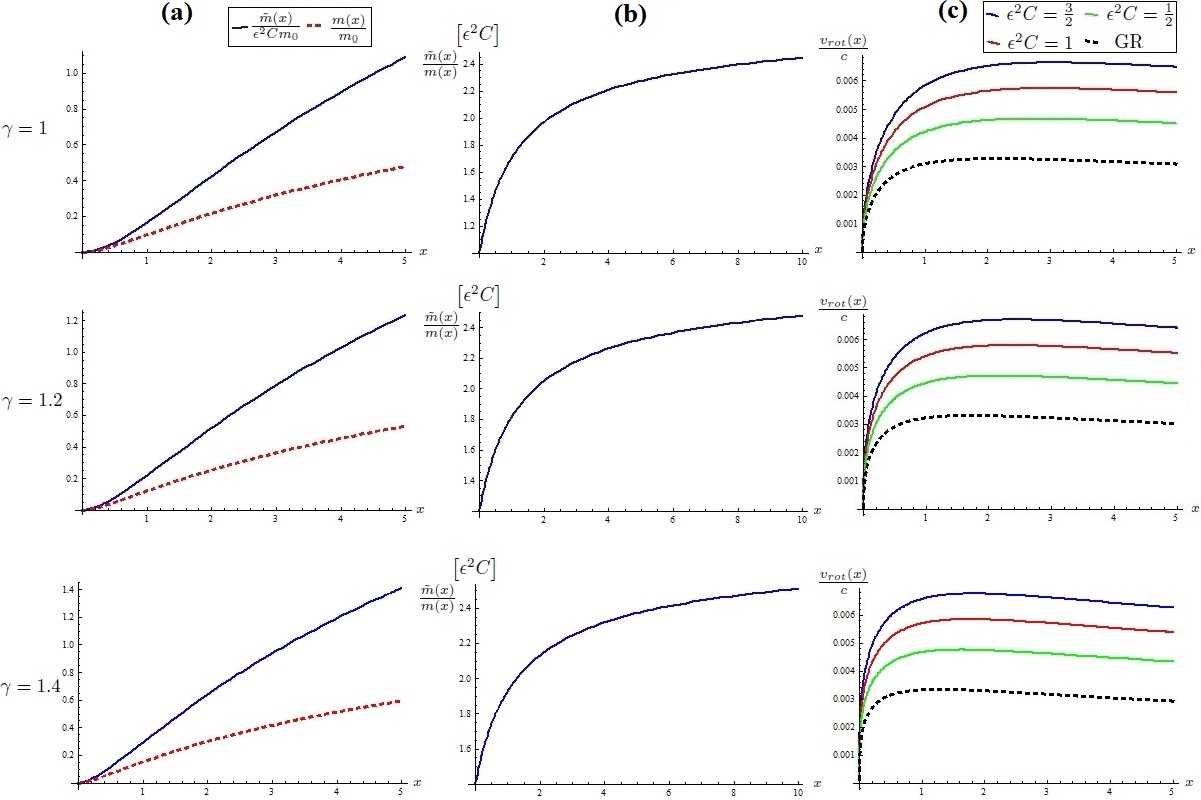}
\caption{{\scriptsize Navarro-Frenk-White Profile Calculation for $\gamma = \{1,1.2,1.4\}$. \textbf{(a)} Ordinary and $\tilde{\delta}$ Matter vs normalize radius, with $m_0 = 8\pi R^3 \rho_0$. We have that more $\tilde{\delta}$ Matter is accumulated from the center, just like the exponential and Einasto cases. \textbf{(b)} $\frac{\tilde{m}(x)}{m(x)}$ (in units of $\epsilon^2C$) vs normalize radius. Comparing with the others profiles, we have that $\tilde{m}(x)$ increases faster than $m(x)$, but in this case the relation between both masses is practically logarithmic. \textbf{(c)} Rotation velocity vs normalize radius for different values of $\epsilon^2C$. The Black-Dashed line correspond to GR case and the others $\epsilon^2C$ values indicate the contribution of $\tilde{\delta}$ Matter. Here we verify the result in the others profiles, $\tilde{\delta}$ Matter amplify the rotation velocity, such that if $\epsilon^2C$ is bigger, we have higher velocities. In these calculations we have used $\kappa \rho_0 R^2 = \frac{\left(3-\gamma\right)^{3-\gamma}}{4\left(2-\gamma\right)^{2-\gamma}} \times 10^{-4}$.}}
\label{Fig: NFW Sol}
\end{center}
\end{figure}

We can also use (\ref{m ord}) and (\ref{m tild}) to obtain the appropriate initial conditions. They are $m(x) \approx \frac{4\pi R^3 \rho_0}{3-\gamma}x^{3-\gamma}$ and $\tilde{m}(x) \approx \frac{4\pi \epsilon^2C R^3 \rho_0\gamma}{3-\gamma}x^{3-\gamma}$, then $\epsilon^2C \sim O(1)$ too. After solving (\ref{sol phi NWF}) and (\ref{sol tphi NFW}), we obtain $m(x)$ and $\tilde{m}(x)$, represented in \textbf{Figure \ref{Fig: NFW Sol}a}, and the relation between $m(x)$ and $\tilde{m}(x)$ in \textbf{Figure \ref{Fig: NFW Sol}b} for $\gamma = \{1,1.2,1.4\}$. Just like the exponential and Einasto cases, more $\tilde{\delta}$ Matter is accumulated far from the center. Comparing with the others profiles, we have that $\tilde{m}(x)$ increases faster than $m(x)$ too, but in this case the relation between both kinds of matter is practically logarithmic, expected in a relation Dark Matter/Baryonic Matter. We know that $\gamma=1$ correspond to just-dark-matter distribution and the others cases with $\gamma>1$ consider a baryonic matter effect \cite{NFW 3}, which means that $\tilde{\delta}$ Gravity could give us a greater value of $\gamma$ than GR in a data simulation, so we obtain less ordinary Dark-Matter. In any case, we obtain the same conclusion; The ordinary matter leads over $\tilde{\delta}$ Matter in the center, but rapidly the second one increases until it is completely dominant. So, $\tilde{\delta}$ Matter is concentrated outside of the galactic nucleus.\\

The rotation velocity is presented in \textbf{Figure \ref{Fig: NFW Sol}c} for $\epsilon^2C = \left\{\frac{1}{2}, 1, \frac{3}{2}\right\}$ and GR. In all our profiles, the rotation velocity is amplified by the $\tilde{\delta}$ Matter effect, such that if $C$ is bigger, we have higher velocities.\\

From exponential, Einasto and NFW profiles, we saw that $\tilde{\delta}$ Matter produces an amplified effect to the ordinary matter, affecting the rotation velocity. Unfortunately, the $\tilde{\delta}$ Matter is principally concentrated outside of the galactic nucleus, as opposed to expected Dark Matter distribution. On the other side, from Einasto and NFW profiles, we observed a logarithmic relation between $m(x)$ and $\tilde{m}(x)$. That is:

\begin{eqnarray}
\tilde{m}(x) \sim m(x)\ln(x).
\end{eqnarray}

In this way, we obtained an additional Dark Matter effect. So, we can divide the ordinary matter in Baryonic and Dark Matter, then we have $\tilde{\delta}$ Baryonic and $\tilde{\delta}$ Dark Matter. In a cosmological level, the Einstein equations just take into account the ordinary matter, and $\tilde{\delta}$ Matter appears in the equation of $\tilde{g}_{\mu \nu}$ (See eqs. (\ref{Einst Eq}) and (\ref{tilde Eq}) respectively or \cite{Paper 1}). All these mean that the quantity of ordinary (\textit{real}) Dark Matter is less than GR. The rest of Dark Matter effect is due to $\tilde{\delta}$ Matter components.\\

Now, if we compare the spherically homogeneous profile with the other ones, we note that the $\tilde{\delta}$ Matter effect is only produced by the distribution of the ordinary matter, the scale is not so important. So, if the Dark Matter is principally explained by $\tilde{\delta}$ Dark Matter, then this effect is only important when the distribution of ordinary matter is strongly dynamic, just like a globular clusters (GCs) for example. This means that we could find some of these GCs where the Dark Matter, $\tilde{\delta}$ Matter in this context, is less than Baryon Matter. Evidences of that have been found in \cite{woDM 1,woDM 2}, where an enormous quantity of Dark Matter is not necessary in the formation of some GCs. This computation will be developed in a future work.\\

A more complete calculation could be developed using a specific profile for Dark Matter, a Einasto Profile with a small $\alpha$ or NFW profile with $\gamma=1$ for example, and other profile for Baryonic Matter. With these considerations, we can isolate the Dark Matter effect from Baryonic contribution, including $\tilde{\delta}$ Matter components. In \cite{Einasto 2} was used a multi-components Einasto profile. Another phenomenon to study with $\tilde{\delta}$ Gravity is the anomalous secular increase of the eccentricity of the orbit of the Moon \cite{Moon}. This effect has not been explained yet, so it is a window for new physics. These computations will be also developed in a future work.\\

\newpage

\section*{Conclusions.}

We have proposed a modified model of gravity, where a new gravitational field $\tilde{g}_{\mu \nu}$ is incorporated. This field transforms correctly under general coordinate transformations and presents a new symmetry called $\tilde{\delta}$ Symmetry. Of course, the new action is invariant under these transformations and it is given by (\ref{grav action}). We call this new gravity model $\tilde{\delta}$ Gravity and a quantum field theory analysis of the model has been developed in \cite{delta gravity}.\\

$\tilde{\delta}$ Gravity has been studied at a classical level in previous works. Particulary, we developed the cosmological case, where we obtained an accelerated expansion of the universe without a cosmological constant \cite{DG DE}-\cite{Paper 1}. In this process, we obtained the classical equations of motion. One of them is the Einstein's equation. This equation is preserved because, in any $\tilde{\delta}$ Theory, the equation of the original fields, as $g_{\mu \nu}$, are not modified. However, we have new equations to solve $\tilde{g}_{\mu \nu}$. Besides, the free trajectory of a massive particle is given by (\ref{geodesics m}). This equation corresponds an anomalous geodesic, so that an effective metric can not be defined. Finally, we fix the gauge for $g_{\mu \nu}$ and $\tilde{g}_{\mu \nu}$, using the extended harmonic gauge, given by (\ref{Harmonic gauge}) and (\ref{Harmonic gauge tilde}). With this basic set up, we can study several phenomena.\\

In this paper, we focused to study the Non-Relativistic limit. In the Newtonian limit, we obtained a similar expression as in GR, where we have an effective potential. This potential depends on $\rho^{(0)}$ and $\tilde{\rho}^{(0)}$, where the last one corresponds to $\tilde{\delta}$ Matter. We found a relation between $\rho^{(0)}$ and $\tilde{\rho}^{(0)}$ and we used it in different density profiles for a galaxy to explain Dark Matter. In first place, we can see that in a spherically homogenous density the $\tilde{\delta}$ Matter effect is completely null. If we consider this profile like a first approximation to a planet or a star, we conclude that in these cases do not have $\tilde{\delta}$ Matter contributions, so it is equivalent to GR. On the other side, with other kind of densities, where the matter distribution changes in the space, we obtain important modifications. For this, we used the exponential, Einasto and NFW profiles, where we noted that $\tilde{\delta}$ Matter produce an amplifying effect in the total mass and rotation velocity in a galaxy. Considering all these, we can say that the $\tilde{\delta}$ Matter effect is only important when the distribution of ordinary matter is strongly dynamic, so the scale is not so important. Therefore, considering $\tilde{\delta}$ Matter as Dark Matter, large structures with a small quantity of Dark Matter should be found \cite{woDM 1,woDM 2}.\\

Finally, we saw that $\tilde{\delta}$ Matter has a special behavior, more similar to Dark Matter compared with its equivalent ordinary component. We can see this in \textbf{Figure \ref{Fig: Einasto Sol} and \ref{Fig: NFW Sol}}, where a logarithmic relation between $\tilde{\delta}$ and Ordinary Matter is observed. In \cite{Paper 1} is computed the $\tilde{\delta}$ Matter quantity in the present at cosmological level. We obtained that the $\tilde{\delta}$ non-relativistic Matter is 23\% the ordinary non-relativistic matter, where Dark Matter is included, implying that Dark Matter is in part $\tilde{\delta}$ Matter. A more complete calculation can be developed if we use a multi-components profile to simulate data of some galaxies \cite{Einasto 2}. In this way, we can isolate the different contributions: Ordinary Baryonic Matter, Ordinary Dark Matter, $\tilde{\delta}$ Baryonic Matter and $\tilde{\delta}$ Dark Matter. Analogous result must be obtained in the CMB Power Spectrum. With all these, we concluded that the Dark Matter effect could be explained with a considerably less quantity of Ordinary Dark Matter, considering that the principal source of this effect is $\tilde{\delta}$ Dark Matter. This result would explain the problematic detection of Dark Matter, however a field theory description of $\tilde{\delta}$ Gravity is necessary to understand the nature of $\tilde{\delta}$ Matter.\\

\newpage

\section*{Appendix A: $\tilde{\delta}$ Theories.}

In this Appendix, we will define the $\tilde{\delta}$ Theories in general and their properties. For more details, see \cite{delta gravity,AppendixA}.\\

\subsection*{$\tilde{\delta}$ Variation:}

These theories consist in the application of a variation represented by $\tilde{\delta}$. As a variation, it will have all the properties of a usual variation such as:

\begin{eqnarray}
\tilde{\delta}(AB)&=&\tilde{\delta}(A)B+A\tilde{\delta}(B) \nonumber \\
\tilde{\delta}\delta A &=&\delta\tilde{\delta}A \nonumber \\
\tilde{\delta}(\Phi_{, \mu})&=&(\tilde{\delta}\Phi)_{, \mu},
\end{eqnarray}

where $\delta$ is another variation. The particular point with this variation is that, when we apply it on a field (function, tensor, etc.), it will give new elements that we define as $\tilde{\delta}$ fields, which is an entirely new independent object from the original, $\tilde{\Phi} = \tilde{\delta}(\Phi)$. We use the convention that a tilde tensor is equal to the $\tilde{\delta}$ transformation of the original tensor when all its indexes are covariant. This means that $\tilde{S}_{\mu \nu \alpha ...} \equiv \tilde{\delta}\left(S_{\mu \nu \alpha ...}\right)$ and we raise and lower indexes using the metric $g_{\mu \nu}$. Therefore:

\begin{eqnarray}
\tilde{\delta}\left(S^{\mu}_{~ \nu \alpha ...}\right)
&=& \tilde{\delta}(g^{\mu \rho}S_{\rho \nu \alpha ...}) \nonumber \\
&=& \tilde{\delta}(g^{\mu \rho})S_{\rho \nu \alpha ...} + g^{\mu \rho}\tilde{\delta}\left(S_{\rho \nu \alpha ...}\right) \nonumber \\
&=& - \tilde{g}^{\mu \rho}S_{\rho \nu \alpha ...} + \tilde{S}^{\mu}_{~ \nu \alpha ...},
\end{eqnarray}

where we used that $\delta(g^{\mu \nu}) = - \delta(g_{\alpha \beta})g^{\mu \alpha}g^{\nu \beta}$.\\

\subsection*{$\tilde{\delta}$ Transformation:}

With the previous notation in mind, we can define how a tilde component transform. In general, we can represent a transformation of a field $\Phi_i$ like:

\begin{eqnarray}
\bar{\delta} \Phi_i = \Lambda_i^j(\Phi) \epsilon_j,
\end{eqnarray}

where $\epsilon_j$ is the parameter of the transformation. Then $\tilde{\Phi}_i = \tilde{\delta}\Phi_i$ transforms:

\begin{eqnarray}
\label{tilde trans general}
\bar{\delta} \tilde{\Phi}_i = \tilde{\Lambda}_i^j(\Phi) \epsilon_j + \Lambda_i^j(\Phi) \tilde{\epsilon}_j,
\end{eqnarray}

where we used that $\tilde{\delta}\bar{\delta} \Phi_i = \bar{\delta}\tilde{\delta} \Phi_i = \bar{\delta}\tilde{\Phi}_i$ and $\tilde{\epsilon}_j = \tilde{\delta} \epsilon_j$ is the parameter of the new transformation. These extended transformations form a close algebra \cite{AppendixA}.\\

Now, we consider general coordinate transformations or diffeomorphism in its infinitesimal form:

\begin{eqnarray}
\label{xi 0}
x'^{\mu} &=& x^{\mu} - \xi_0^{\mu}(x) \nonumber \\
\bar{\delta} x^{\mu} &=& - \xi_0^{\mu}(x),
\end{eqnarray}

where $\bar{\delta}$ will be the general coordinate transformation from now on. Defining:

\begin{eqnarray}
\label{xi 1}
\xi_1^{\mu}(x) \equiv \tilde{\delta} \xi_0^{\mu}(x)
\end{eqnarray}

and using (\ref{tilde trans general}), we can see a few examples of how some elements transform:\\

\textbf{I)} A scalar $\phi$:

\begin{eqnarray}
\label{scalar}
\bar{\delta} \phi &=& \xi^{\mu}_0 \phi_{, \mu} \\
\label{scalar_tild}
\bar{\delta} \tilde{\phi} &=& \xi^{\mu}_1 \phi,_{\mu} + \xi^{\mu}_0 \tilde{\phi},_{\mu}.
\end{eqnarray}

\textbf{II)} A vector $V_{\mu}$:

\begin{eqnarray}
\label{vector}
\bar{\delta} V_{\mu} &=& \xi_0^{\beta} V_{\mu, \beta} + \xi_{0, \mu}^{\alpha} V_{\alpha} \\
\label{vector_tild}
\bar{\delta} \tilde{V}_{\mu} &=& \xi_1^{\beta} V_{\mu, \beta} + \xi_{1, \mu}^{\alpha} V_{\alpha} + \xi_0^{\beta} \tilde{V}_{\mu, \beta} + \xi_{0, \mu}^{\alpha} \tilde{V}_{\alpha}.
\end{eqnarray}

\textbf{III)} Rank two Covariant Tensor $M_{\mu \nu}$:

\begin{eqnarray}
\label{tensor}
\bar{\delta} M_{\mu \nu} &=& \xi^{\rho}_0 M_{\mu \nu, \rho} + \xi_{0,\nu}^{\beta} M_{\mu \beta} + \xi_{0,\mu}^{\beta} M_{\nu \beta} \\
\label{tensor_tild}
\bar{\delta} \tilde{M}_{\mu \nu}  &=& \xi^{\rho}_1 M_{\mu \nu, \rho} + \xi_{1, \nu}^{\beta} M_{\mu \beta} + \xi_{1, \mu}^{\beta} M_{\nu \beta} + \xi^{\rho}_0 \tilde{M}_{\mu \nu, \rho} + \xi_{0, \nu}^{\beta} \tilde{M}_{\mu \beta} + \xi_{0, \mu}^{\beta} \tilde{M}_{\nu \beta}.
\end{eqnarray}

These new transformations are the basis of $\tilde{\delta}$ Theories. Particulary, in gravitation we have a model with two fields. The first one is just the usual gravitational field $g_{\mu \nu}$ and the second one is $\tilde{g}_{\mu \nu}$. Then, we will have two gauge transformations associated to general coordinate transformation. We will call it extended general coordinate transformation, given by:

\begin{eqnarray}
\label{trans g}
\bar{\delta} g_{\mu \nu} &=& \xi_{0 \mu ; \nu} + \xi_{0 \nu ; \mu} \\
\label{trans gt}
\bar{\delta} \tilde{g}_{\mu \nu} ( x ) &=& \xi_{1 \mu ; \nu} + \xi_{1\nu ; \mu} + \tilde{g}_{\mu \rho} \xi_{0, \nu}^{\rho} + \tilde{g}_{\nu \rho} \xi^{\rho}_{0, \mu} + \tilde{g}_{\mu \nu,\rho} \xi_0^{\rho},
\end{eqnarray}

where we used (\ref{tensor}) and (\ref{tensor_tild}). Now, we can introduce the $\tilde{\delta}$ Theories.\\

\subsection*{Modified Action:}

In the last section, the extended general coordinate transformations were defined. So, we can look for an invariant action. We start by considering a model which is based on a given action $S_0[\phi_I]$ where $\phi_I$ are generic fields, then we add to it a piece which is equal to a $\tilde{\delta}$ variation with respect to the fields and we let $\tilde{\delta} \phi_J = \tilde{\phi}_J$, so that we have:

\begin{eqnarray}
\label{Action}
S [\phi, \tilde{\phi}] = S_0 [\phi] +  \int d^4x \frac{\delta S_0}{\delta \phi_I(x)}[\phi] \tilde{\phi}_I(x),
\end{eqnarray}

the index $I$ can represent any kinds of indices. (\ref{Action}) give us the basic structure to define any modified element for $\tilde{\delta}$ type theories. In fact, this action is invariant under our extended general coordinate transformations developed previously. For this, see \cite{AppendixA}.\\

A first important property of this action is that the classical equations of the original fields are preserved. We can see this when (\ref{Action}) is varied with respect to $\tilde{\phi}_I$:

\begin{eqnarray}
\label{Eq_phi}
\frac{\delta S_0}{\delta \phi_I(x)}[\phi] = 0.
\end{eqnarray}

Obviously, we have new equations when varied with respect to $\phi_I$. These equations determine $\tilde{\phi}_I$ and they can be reduced to:

\begin{eqnarray}
\label{Eq_phi_tilde}
\int d^4x \frac{\delta^2 S_0}{\delta \phi_I(y) \delta \phi_J(x)}[\phi] \tilde{\phi}_J(x) = 0.
\end{eqnarray}

\newpage
\section*{Appendix B: Test Particle Action.}

To understand how the new fields affect the trajectory of a particle, we need to study the Test Particle Action. The first discussion of this issue in $\tilde{\delta}$ Gravity is in \cite{DG DE}. In GR, the action for a test particle is given by:

\begin{eqnarray}
\label{Geo Action 0}
S_0[\dot{x},g] = - m \int dt \sqrt{-g_{\mu \nu}\dot{x}^{\mu}\dot{x}^{\nu}},
\end{eqnarray}

with $\dot{x}^{\mu} = \frac{d x^{\mu}}{d t}$. This action is invariant under reparametrizations, $t' = t - \epsilon(t)$. In the infinitesimal form is:

\begin{eqnarray}
\label{reparametr}
\delta_R x^{\mu} &=& \dot{x}^{\mu}\epsilon.
\end{eqnarray}

In $\tilde{\delta}$ Gravity, the action is always modified using (\ref{Action}) from \textbf{Appendix A}. So, applying it to (\ref{Geo Action 0}), the new test particle action is:

\begin{eqnarray}
\label{Geo Action 01}
S[\dot{x},y,g,\tilde{g}] &=& - m \int dt \sqrt{-g_{\mu \nu}\dot{x}^{\mu}\dot{x}^{\nu}} + \frac{m}{2} \int dt \left(\frac{\tilde{g}_{\mu \nu}\dot{x}^{\mu}\dot{x}^{\nu}+g_{\mu \nu,\rho}y^{\rho}\dot{x}^{\mu}\dot{x}^{\nu}+2g_{\mu \nu}\dot{x}^{\mu}\dot{y}^{\nu}}{\sqrt{-g_{\mu \nu}\dot{x}^{\mu}\dot{x}^{\nu}}}\right) \nonumber \\
&=& m \int dt \left(\frac{\left(g_{\mu \nu} + \frac{1}{2}\tilde{g}_{\mu \nu}\right)\dot{x}^{\mu}\dot{x}^{\nu} + \frac{1}{2}(2g_{\mu \nu}\dot{y}^{\mu}\dot{x}^{\nu} + g_{\mu \nu, \rho}y^{\rho}\dot{x}^{\mu}\dot{x}^{\nu})}{\sqrt{-g_{\alpha \beta}\dot{x}^{\alpha}\dot{x}^{\beta}}}\right),
\end{eqnarray}

where we have defined $y^{\mu} = \tilde{\delta}x^{\mu}$ and we used that $g_{\mu \nu} = g_{\mu \nu}(x)$, so $\tilde{\delta}g_{\mu \nu} = \tilde{g}_{\mu \nu} + g_{\mu \nu, \rho}y^{\rho}$. Naturally, this action is invariant under reparametrization transformations, given by (\ref{reparametr}), plus $\tilde{\delta}$ reparametrization transformations:

\begin{eqnarray}
\label{reparametr plus}
\delta_R y^{\mu} &=& \dot{y}^{\mu}\epsilon + \dot{x}^{\mu}\tilde{\epsilon},
\end{eqnarray}

just like it is shown in (\ref{tilde trans general}). The presence of $y^{\mu}$ suggests additional coordinates, but our model just live in four dimensions, given by $x^{\mu}$. Actually, $y^{\mu}$ can be gauged away using the extra symmetry corresponding to $\tilde{\epsilon}$ in equation (\ref{reparametr plus}), imposing the gauge condition $2g_{\mu \nu}\dot{y}^{\mu}\dot{x}^{\nu} + g_{\mu \nu, \rho}y^{\rho}\dot{x}^{\mu}\dot{x}^{\nu} = 0$. However, the extended general coordinate transformations as well as the usual reparametrizations, given by (\ref{reparametr}), are still preserved. Then, (\ref{Geo Action 01}) can be reduced to:

\begin{eqnarray}
\label{Geo Action}
S[\dot{x},g,\tilde{g}] = m \int dt \left(\frac{\left(g_{\mu \nu} + \frac{1}{2}\tilde{g}_{\mu \nu}\right)\dot{x}^{\mu}\dot{x}^{\nu}}{\sqrt{-g_{\alpha \beta}\dot{x}^{\alpha}\dot{x}^{\beta}}}\right).
\end{eqnarray}

Notice that the test particle action in Minkowski space is recovered if we used the boundary conditions, given by $g_{\mu \nu} \sim \eta_{\mu \nu}$ and $\tilde{g}_{\mu \nu} \sim 0$, to be equal to GR. So, this action for a test particle in a gravitational field will be considered as the starting point for the physical interpretation of the geometry in $\tilde{\delta}$ Gravity. Now, the trajectory of massive test particles is given by the equation of motion of $x^{\mu}$. This equation say us that $g_{\mu \nu}\dot{x}^{\mu}\dot{x}^{\nu} = cte$, just like GR. Now, if we choose $t$ equal to the proper time, then $g_{\mu \nu}\dot{x}^{\mu}\dot{x}^{\nu} = -1$  and the equation of motion is reduced in this case to:

\begin{eqnarray}
\label{geodesics m}
\hat{g}_{\mu \nu} \ddot{x}^{\nu} + \hat{\Gamma}_{\mu \alpha \beta} \dot{x}^{\alpha} \dot{x}^{\beta} = \frac{1}{4}\tilde{K}_{,\mu},
\end{eqnarray}

with:

\begin{eqnarray}
\hat{\Gamma}_{\mu \alpha \beta} &=& \frac{1}{2}(\hat{g}_{\mu \alpha , \beta} + \hat{g}_{\beta \mu , \alpha} - \hat{g}_{\alpha \beta ,
\mu})\nonumber \\
\hat{g}_{\alpha \beta} &=& \left(1+\frac{1}{2} \tilde{K}\right)g_{\alpha \beta} + \tilde{g}_{\alpha \beta}
\nonumber \\
\tilde{K} &=& \tilde{g}_{\alpha \beta} \dot{x}^{\alpha} \dot{x}^{\beta}. \nonumber
\end{eqnarray}

This equation of motion is independent of the mass of the particle, so all particles will fall with the same acceleration. On the other side, the equation (\ref{geodesics m}) is a second order equation, but it is not a classical geodesic, because we have additional terms and an effective metric can not be defined. On the other side, it is important to observe that the proper time is defined in the same way than GR, because the equation of motion satisfies that the quantity that is constant is $g_{\mu \nu}\dot{x}^{\mu}\dot{x}^{\nu} = -1$. So, we must define proper time using the original metric $g_{\mu \nu}$. However, (\ref{geodesics m}) is useless for massless particles, because (\ref{Geo Action 0}) is null when $m=0$, and this is important to define the three-dimensional metric. We can prove that it is determined by both tensor fields, $g_{\mu \nu}$ and $\tilde{g}_{\mu \nu}$, so the geometry is affected by $\tilde{\delta}$ components. More details of that are detailed on \cite{Paper 1}.


\section*{Acknowledgements.}

The work of P. Gonz\'alez has been partially financed by Beca Doctoral Conicyt $N^0$ 21080490, Fondecyt 1110378, Anillo ACT 1102, Anillo ACT 1122 and CONICYT Programa de Postdoctorado FONDECYT $N^o$ 3150398. The work of J. Alfaro is partially supported by Fondecyt 1110378, Fondecyt 1150390, Anillo ACT 1102 and Anillo ACT 11016. J.A. wants to thank F. Prada and R. Wojtak for useful remarks.\\


\end{document}